\documentclass{aastex}

\usepackage{graphicx,shapepar}

\newcommand{\kms}{\ensuremath{\mathrm{km\ s}^{-1}}}

\newcommand{\nii}{[\ion{N}{2}]}

\newcommand\NIIlam{[\ion{N}{2}]\,6584\,}

\newcommand\Ha{\ensuremath{\mathrm{H}\alpha}}

\newcommand{\vhel}{\ensuremath{V_\mathrm{hel}}}

\newcommand{\vsys}{\ensuremath{V_\mathrm{sys}}}
\newcommand{\vexp}{\ensuremath{V_\mathrm{exp}}}

\newcommand{\teff}{\ensuremath{T_\mathrm{eff}}}

\title{A DETAILED MORPHO-KINEMATIC MODEL OF THE ESKIMO, NGC 2392. A UNIFYING VIEW WITH THE CAT'S EYE  AND SATURN PLANETARY  NEBULAE}

\author{Ma.\ T. Garc\'{\i}a-D\'{\i}az, J. A. L\'opez, W. Steffen \& 
  M. G., Richer 
 \affil{Instituto de Astronom\'ia,
    Universidad Nacional Aut\'onoma de M\'exico} Campus Ensenada,
  Ensenada, Baja California, 22800, M\'exico}
\affil{tere@astrosen.unam.mx, jal@astrosen.unam.mx, wsteffen@astrosen.unam.mx, richer@astrosen.unam.mx}

\begin{abstract}
The 3--D and kinematic structure of the Eskimo nebula, NGC~2392, has
been notoriously difficult to interpret in detail given its complex
morphology, multiple kinematic components and its nearly pole--on
orientation along the line of sight. We present a comprehensive,
spatially resolved, high resolution, long-slit spectroscopic mapping
of the Eskimo planetary nebula. The data consist of 21 spatially
resolved, long--slit echelle spectra tightly spaced over the Eskimo
and along its bipolar jets. This data set allows us to construct a
velocity--resolved [NII] channel map of the nebula with a resolution
of 10~\kms{} that disentangles the different kinematic components of
the nebula. The spectroscopic information is combined with {\it HST}
images to construct a detailed three dimensional morpho--kinematic
model of the Eskimo using the code SHAPE. With this model we
demonstrate that the Eskimo is a close analog to the Saturn and the
Cat's Eye nebulae, but rotated 90\degr{} to the line of sight.
Furthermore, we show that the main characteristics of our model apply
to the general properties of the group of elliptical planetary nebulae
with ansae or FLIERS, once the orientation is considered. We conclude
that these kind of nebulae belongs to a class with a complex common
evolutionary sequence of events.

\end{abstract}

\keywords{Planetary Nebulae: individual (NGC~2392, NGC~7009, NGC~6543)
  $-$ ISM: kinematics and dynamics $-$ ISM jets and outflows
  $-$ techniques: spectroscopy}

\begin{document}
\maketitle

\section{INTRODUCTION}
\label{sec:introduction}

Planetary nebulae (PNe) are formed by low to intermediate mass
stars. At the end of their lives, during the thermally pulsating
asymptotic giant branch (AGB) stage, these stars expel their
chemically enriched outer layers and begin a rapid transformation of
their degenerate carbon--oxygen core towards their final evolution as
a white dwarf. Along this route, the previously expelled, cold,
neutral, molecular and dusty shell is fully ionized by the progenitor
star in only several hundred to a few thousand years while the system
suffers a transition stage known as planetary nebula. Once the central
star (CS) leaves the AGB the stellar wind gradually develops
supersonic velocities and shocks the freshly ionized shell,
contributing important thermal and hydrodynamic effects to the shell
development. During this process the nebula also suffers constant
changes in terms of optical depth of the expanding shell, ionizing
conditions and evolutionary shaping takes place.  All these changes
are reflected in the varying spectral emission conditions and
morphology of the PN as this evolves.

The Eskimo is a moderately evolved, high excitation PN that presents a
roundish outline on the sky, although it is well known that this
appearance is only due to its polar orientation towards the observer
since the real morphology of its main shell approximates a prolate
spheroidal (e.g. Weedman 1968). \citet{Odell85} and
\citet{Gieseking85} discovered the first bipolar jet in a PN in the
Eskimo; a highly collimated, high-speed outflow directed nearly
towards (and away from) the observer. The bipolar jet is not detected
in optical images due to its particular orientation and it is only
apparent in high resolution, spatially resolved spectra. In addition
to the former authors, conceptual models of the Eskimo based on high
resolution spectroscopy and narrow-band images have been presented by
\citet{Reay83}, \citet{Balick87c}, \citet{Odell90} and \citet{Phillips99}, and \citet{Dufour12}. Although
with some differences in their models, they all consider a fast
expanding, prolate spheroidal inner shell, an outer disk and an outer
envelope, the fast bipolar outflow and the knotty structure in the fur
of the Eskimo.  However, \citet{Zhang12} have recently argued that NGC
2392 is a twin to the tight-waist bipolar planetary Mz-3, but they
built their model mostly on arguments rather than data, and as we will
show here their model does not agree with the most salient features
present in our data.

NGC 2392 is also a point and diffuse X-ray source \citep{Guerrero05, Kastner12}, the latter produced by a wind-shock generated hot bubble and likely related to the fast expansion of the inner shell. The origin of the X-ray point source is less clear, the central star (CS) of the Eskimo has been studied by several
authors.  \citet{Heap77} classified the CS in the Eskimo as spectral type O6f with
\teff{} $=~35~000$~K, a result in conflict with that derived from
 the He~II Zanstra temperature that yields $\approx$~92~000~K. This discrepancy 
 lead Heap to suggest that there must be a second hotter star
responsible for the nebular ionization.  Several authors have calculated an effective
temperature, \teff, in the range 40 000 K -- 45 000 K.
\citep{Mendez11, Pauldrach04, Kudritzki97}. However this temperature
is not high enough to explain the high stages of ionization from some
ions such as O~IV and Ne~V which have been observed in the nebula
\citep{Pottasch08, Natta80}. Other studies have suggested that the effective 
temperature of the CS must be around
74 000 to 80 000 K \citep{Tinkler02, Pottasch08}.
\citet{Pottasch08} find that the central star must have evolved from a
1.7~M$_{\odot}$ progenitor.  
  \citet{Ciardullo99} detected a possible
faint companion to the CS with {\it HST} in the I band but invisible in the V band. 
Recently,  \citet{Danehkar12} used photoionization
models to estimate that the companion star must have a
\teff{}~=~250 000 K, which is much higher than any other value proposed earlier.
Although the presence of a faint and hot companion to the CS of the Eskimo is likely,
there is no firm or unambiguous evidence of its existence to date.

The most recent distance estimates to the Eskimo nebula are those of   \citet{Stanghellini08} with d = 1.6~kpc$\pm130$ pc, \citet{Frew08} with d = 1.28 kpc $\pm130$ pc and \citet{Pottasch11} with d = 1.8 kpc.  

The spectroscopic observations reported here were obtained with the aim to carefully disentangle
the different kinematic components of the Eskimo nebula and their relative location in space . We use the morpho--kinematical modeling code SHAPE to construct  a detailed three dimensional model of the Eskimo from  the kinematic data  combined with  high spatial resolution images from {\it HST}. A 3D model is produced as the final output that clearly illustrates the structure of the Eskimo nebula. In addition, we identify significant structural similarities between the Eskimo (NGC~ 2392), the Saturn (NGC~7009) and the Cat's Eye (NGC~6543) nebulae, and in general with the family of elliptical PNe with ansae or FLIERS, that provide a fresh unifying view of this group of PNe.
 In \S\S~2$-$3 we first describe the observations,
data reduction, and the
various morpho--kinematic structures of the nebula. Isovelocity channel maps are presented in \S4. Then we
describe the 3--D SHAPE model of NGC~2392 and a comparison of the
Eskimo with the Saturn and Cat's Eye nebula (\S 5). Finally in \S6 we
summarize the conclusions of this study.

\section{OBSERVATIONS AND DATA REDUCTION}
\label{sec:observations}

Long-slit, echelle, spectroscopic observations of NGC 2392 were obtained at the
Observatorio Astron\'omico Nacional at  San Pedro M\'artir, (SPM), Baja California,
M\'exico, with the Manchester Echelle Spectrometer
(MES-SPM) \citep{Meaburn03} on the 2.1 m telescope in its $f$/7.5
configuration. The observing run took place in 2002, January 7 -- 10.
MES-SPM was equipped with a SITE-3 CCD
detector  with 1024 $\times$ 1024
square pixels, each 24 $\mu$m on a side ($\equiv$ 0.312 arcsec pixel$^{-1}$).  We used a
90 \AA{} bandwidth filter to isolate the 87th order containing the
\Ha{} and \nii{} nebular emission lines. 
Two pixel binning was employed in both the spatial and spectral
directions. Consequently, 512 increments, each 0\farcs624{} long gave
a projected slit length of 5\farcm32 on the sky, except for slit
position g where we used no binning (i.e. $1024 \times 1024$). 
For the majority of the exposures, we
used a 70~$\mu$m{} ($\equiv$0.95\arcsec) slit, giving a velocity
resolution of 9.2 \kms{} ($\equiv$ 0.312 arcsec pixel$^{-1}$ and 5.6 \kms{} for slit g) and for two exposures, slits a \& r, we used
150 $\mu$m{} ($\equiv$ 1\farcs 9 and $\equiv$ 11.5 \kms). We obtained 18 consecutive and tightly spaced positions over the
Eskimo with  P.A.~=~0\degr,  one position with 
P.A.~=~110\degr{} across the center of the nebula, slit s, and two
positions with  P.A.~=~70\degr{}, slits t \& u. Slit u sits on the bipolar jet. The
slit positions are indicated and
labeled in Figure 1 on a WFPC~2 image of the Eskimo obtained from the HST archive.
All spectra were acquired using exposure times of 1800~s. In
order to establish the exact position of the slit in each pointing, the slit position on the sky was recorded with an automatic procedure available in MES-SPM prior to the spectroscopic exposure. 


The data was reduced by using standard IRAF\footnote{IRAF is
  distributed by the National Optical Astronomy Observatories, which
  is operated by the Association of Universities for Research in
  Astronomy, Inc. under cooperative agreement with the National
  Science foundation} tasks to correct bias and remove cosmic rays. The
spectra were calibrated in wavelength against the spectrum of a Th/Ar
lamp to an accuracy of $\pm$1 \kms{} when converted to radial
velocity.  All spectra presented in this paper are corrected to
heliocentric velocity (\vhel).

 All the spectroscopic data have been drawn from 
 {\it ``The San Pedro M\'artir Kinematic Catalogue of
  Galactic Planetary Nebulae''} \citep{Lopez12} and  available at
  http://kincatpn.astrosen.unam.mx. In the last part of this work we also make use 
  of spectra for the Saturn (NGC 7009) and Cat's Eye (NGC 6543) nebulae, these have
  also been drawn from the same source, they both have been observed with the same detector
  as described above, binned $2 \times 2$ and with a 150~$\mu$m slit ($\equiv$ 1\farcs 9 and $\equiv$ 11.5 \kms).
The spectra for the Saturn and Cat's Eye nebulae are shown
in Figure 9.

Additional steps where taken to produce velocity cubes in
\NIIlam. These were implemented using FORTRAN routines. The additional
tasks were: a) Continuum emission was removed by fitting a quadratic
function to each row of each two-dimensional spectrum. b) An astrometric
solution was employed for each of the image+slit using several stars
near the Eskimo in order to find the exact slit position of each
exposure. The position of each star was taken from Sky View data. c)
The data were photometrically calibrated using the \NIIlam{} HST
image. In this way, the observed slit spectra oriented N-S were
combined and interpolated to constructed isovelocity channel maps of
the Eskimo. This techniques are described in detail in
\citet{Garcia07}.

The resulting data cubes for the \NIIlam{} line are shown in
Figure 5 as isovelocity channel maps, each isovelocity map is 60 \kms{} wide (see \citet{Balick87a} for a similar data representation with narrower velocity intervals.). The derived moment maps for the same emission line are shown in Figure 6. 

\section{KINEMATICS}
\label{sec:kinematic}

The bi-dimensional \nii{} emission
line spectra or position--velocity ({\it  P--V}) arrays for all individual slit position are shown in
Figures~2 and 3 (except for slit position r  since the  \nii{} line emission was extremely faint at this location).  For each slit position we show a couple of {\it  P--V} arrays: the observed \nii{} {\it  P--V} array is on the left panel and on the right panel is the corresponding synthetic {\it  P--V} array derived from the SHAPE model. Spatial offsets are in arcseconds. The stellar continuum from the CS has not been subtracted and this is apparent in the {\it  P--V} arrays corresponding to slit positions j, s and u.
 All spectra are presented within a heliocentric
velocity range of $-$135 to $+$278~\kms. We choose this range because
it allows to appreciate the full complex structure of the line profiles and
clearly identifies the beginning and the end of the high-velocity, bipolar, collimated
outflow.  We derive a heliocentric systemic
velocity, \vsys $= 70.5$~\kms{} by using slit position j, which passes
through the central star.

In order to facilitate the discussion of the kinematic information, we have
labeled in Figure 4 each emission region seen in  the {\it  P--V} arrays in slit positions j and u.
Each of these components are present throughout the spectra presented in Figures 2 and 3 
and are the basis of the interpretation and our model of the Eskimo. A description of each component is given below, please refer to Figure 4 for the remainder of this section.

\subsection{THE INNER SHELL}

The inner shell is recorded from slit positions g to m that cut the nebula at P.A. = 0\degr{}  over the central region and also in slits u and s (P.A. = 70\degr). The line profiles from the inner shell corresponds to a tilted and distorted (peanut-like) velocity ellipse ({\it VE}) with bright regions over its outline that are part of the filamentary structure projected over the surface of this shell. In the central slit, j, the velocity ellipse covers a velocity range   \vhel{} $\simeq -$50 to 190~\kms{}, considering that \vsys $= 70.5$~\kms{}, then the expansion velocity of the inner shell along the line of sight is \vexp  $\simeq \pm 120$ ~\kms{}, this is probably the largest expansion velocity for an elliptical shell in a PN and presumably corresponds to the expansion of the prolate ellipsoid, indicated by the shape of the {\it  P--V} array, close to its major axis. Note that the tilt of the {\it VE} is more pronounced in slit g than in slit m, that correspond to the edges of the shell. There is also a small velocity shift in the limits of the {\it VE's} corresponding to the contiguous positions on either side of the central slit. This shift cannot be read out from the {\it P--V} arrays in Figure 2 due to the scale, this shift is of the order of 10 ~\kms{};  for slit i the {\it VE} is slightly redshifted and for slit k the shift is in the opposite sense. These elements indicate that the main axis of the fast-expanding  ellipsoidal shell is pointing not directly towards the observer (pole on) but slightly south and to the southwest. This is in accordance with the model of  \citet{Odell90} that derived a P. A. = 202\degr{} for the main axis of the ellipsoid or prolate spheroidal shell and an inclination $\sim -19$\degr{} with respect to the line of sight. In our SHAPE model (see below, section 5) we obtain P. A. = 205\degr{} and an inclination $\sim -9$\degr{} with respect to the line of sight. 

\subsection{THE OUTER SHELL AND COMETARY KNOTS}

The outer shell is revealed throughout all the {\it P--V} arrays in Figures 2 and 3 as the central and narrow velocity ellipse expanding with a velocity  \vexp{} $\simeq 16$ \kms{}.  This shell is usually referred to as the hood of the Eskimo. The shape of the {\it VE's} is smooth and uniform, indicating that the outer shell is close to an expanding sphere. Some of the cometary knots that are seen in projection against the hood of the Eskimo are apparent as bright knots on the extremes of the narrow {\it VE} of the outer shell throughout the different positions where the slits intersects the knots or their bright tails, and their velocities coincide with \vsys, which means that either these knots are inert or any motion in the system of these knots must be tangential to the sightline and implying that they must be distributed in a disk-like configuration. Since the major axis of the inner ellipsoidal shell is nearly pole-on that means that the disk must be close to the plane of the sky and likely coinciding with the equatorial region of inner shell.

\subsection{THE CAPS }

In addition to the inert cometary knots there is an extended system of bright knots and diffuse material located in the outer shell that make-up the fur in the Eskimo's hood. 
These knots are located below and above the {\it VE} that describes the inner shell and shifted in velocity with respect to the thin {\it VE} that represents the outer shell. We call these systems or groups of knots simply caps. Their projected location corresponds to the fur of the Eskimo's hood, as the cometary knots described above, but their radial velocities indicate that they are distributed in three extended groups. The knots located in the northern section  all show only receding velocities whereas the group in the southern section show both receding and  approaching radial velocities but with a different spatial distribution. Those with receding radial velocities appear at projected locations slightly above the knots with blueshifted radial velocities. Considering that the tilt for the inner shell described above applies also to the rest of the fur structure, this projected spatial segregation of the receding and approaching  system of bright knots in the southern section indicates that we are looking at  caps in front and behind the inner shell and with an expansion velocity $\sim \pm 55$ \kms. In the northern section there is only the receding cap and presumably located in the back part of the nebula. The location and approximate extent of the caps is indicated in Figure 4. These caps very much resemble the system of FLIERS and low ionization emission regions,  like those in NGC 7662, NGC 7009 and NGC 6543 \citep[e.g][]{Balick87b, Balick98, Reed99}, see Figure  9.

\subsection{THE COLLIMATED, HIGH VELOCITY, BIPOLAR OUTFLOWS}

The bipolar jets are clearly appreciated in slit u. They are seen to be launched right from the star (the bipolar jets emerge from the stellar continuum)  at $\sim \vhel -95$ and $+235$ \kms;  considering that \vsys{} $= 70.5$ \kms, they are launched from the source at \vexp{} $=\pm 165$ \kms. The jets continue increasing their velocity with distance to reach top velocities \vhel {}$-110$ and $+250$ \kms or \vexp$\simeq{} \pm180$ \kms. At the point where the bipolar jets seem to emerge from the inner shell, approximately 12 arcseconds  away from the stellar continuum, the jet emission suffers a discontinuity and its full extent gets split in two sections, in both directions, at this point both jets seem to suffer a slight acceleration. 
All jet components have similar brightness as well as spatial and velocity extents. The discontinuity in the bipolar jets may arise from a sudden change of environment, when moving from the inner hot shell into the realm of the outer shell or to an episodic ejection.The presence of the bipolar jets is also apparent from slits c - q  at the locations where the slits intersect the jet. The redshifted fast velocity components are observed from slits c to i and the blue-shifted counterparts from slits k to q. The projected width of the jets in these slits is of the order of 10 arcseconds and remains remarkably constant along the entire length of the jet.

\section{ISOVELOCITY CHANNEL MAPS}

In order to provide an alternative overview of the kinematic structure
 of the Eskimo we constructed isovelocity channel maps of
the \nii{} optical emission line. The channel map representations are very useful 
to reveal large-scale, spatially coherent features that are
not readily apparent in the individual long-slit spectra. The maps were
derived from slits a~--~r, each velocity cube is an image of the
nebula within a 60 \kms{} range in velocity, extending from $+$240 to
$-$110 \kms{}. These isovelocity channel maps are presented in Figure 5, where relative
positions are given in arcseconds with respect to the CS. The velocity interval and limits have been chosen to produce a clear snapshot of the main kinematic structures of NGC 2392. The first (a) and last (e) panels in the figure clearly show the collimated bipolar outflow at the extreme velocities covered by the isovelocity channel maps. The second (b) and fourth (d) panels reveal the location of the redshifted and blueshifted caps outside the inner shell, note the slight but apparent spatial displacement of the inner shell between the redshifted (b panel )  and the  blueshifted (d panel) maps due to its tilt with respect to the line of sight. Finally,  the central (c) panel shows the structure of the Eskimo at velocities close to systemic; cometary knots and caps are clearly discerned comparing the central panels.

We have also constructed the kinematic moment maps for the same  \nii{}  
emission line. These maps of derived parameters offer a complementary approach to the isovelocity channel maps. The parameters shown in Figure 6 are the total line surface brightness, $SB$,  mean heliocentric velocity,  $\overline{V_{\textrm{\scriptsize hel}}}$, root-mean-square (RMS) velocity, $\sigma$, and skewness $S_k$. These quantities are defined in terms of the  velocity moments, $M_k$ of the line profiles $I(v)$:

\begin{equation}
\textrm M_k = \int_{v_1}^{v_2}v^k I(v)dv
\label{eq:mom0}
\end{equation}
 
 and the limits of the integration are  $v_1$ = $-$140 \kms{} and $v_2$ = $+$280 \kms, which represent a wide enough range to include most of the emission of interest.
Thus, the zero moment, $M_0$,

\begin{equation}
\textrm{$SB$} = M_0 = \int_{v_1}^{v_2} I(v)dv
\label{eq:mom0}
\end{equation}

  The first velocity moment, $M_1$ is,

\begin{equation}
M_1 = \int_{v_1}^{v_2} v I(v)dv,
\label{eq:mom1}
\end{equation}

and the mean heliocentric velocity is calculated then from the zero and first
moment,

\begin{equation}
 \overline{V_{\textrm{\scriptsize hel}}} = \frac{M_1}{M_0}
\label{eq:mom3}
\end{equation}

The root-mean-square velocity width, $\sigma$, is derived
from the second moment, $M_2$,

\begin{equation}
M_2 = \int_{v_1}^{v_2} v^2 I(v)dv 
\label{eq:mom2}
\end{equation}

then,

\begin{equation}
 \sigma^2 = \frac{M_2}{M_0} - \left(  \overline{V_{\textrm{\scriptsize hel}}} \right)^2
\label{eq:mom4}
\end{equation}

Finally, from the third moment  the velocity skewness map is derived:

\begin{equation}
M_3 = \int_{v_1}^{v_2} v^3 I(v)dv 
\label{eq:mom2}
\end{equation}

and,

\begin{equation}
S_k = M_3 / M_0 
\label{eq:mom2}
\end{equation}

The first panel in Figure 6  shows the map of the \nii{} line intensity distribution or spatial variations over the nebula within the full velocity range of integration. The second panel displays the distribution of $ \overline{V_{\textrm{\scriptsize hel}}} $ that highlights the regions corresponding to the shells and caps. The RMS velocity is displayed in the third panel, since $\sigma$ is sensitives to low emission gas at high velocities , the bipolar collimated outflows become apparent here, finally the last panel contains the skewness map that shows that the velocity field within the nebula is mostly symmetric.

\section{MORPHO-KINEMATIC MODELING WITH SHAPE}
\label{sec:shape}

The 3-D morpho-kinematic structure in the \NIIlam{} emission of NGC 2392 has been modeled using the code SHAPE, developed by \citet{shapeWo06} and \citet{shapeWo10}. Similar analyses have been performed recently on the planetary nebulae NGC 6337 \citep{gd09}, NGC 7009 \citep{Steffen09}, NGC 6751 \citep{clark10},  and Hb 5 \citep{lophb5}.

Modeling with SHAPE follows three main steps. First, defining the geometrical forms to use;
SHAPE has a variety of objects such as a sphere, torus, cone, cube, etc. whose basic forms can be  modified by the user (e.g. squeeze, twist, boolean, etc). Second, an emissivity distribution is assigned to each individual object or structure, and third, a
velocity law is chosen as a function of position. SHAPE gives as
result a two dimensional image and synthetic {\it P-V} arrays, which are
rendered from the 3D model to be compared with the observed data. The parameters of the model are then iteratively adjusted until a satisfactory solution is obtained. 

The Eskimo has been modeled using the \nii{} {\it HST} image and the  {\it  P--V} spectra shown in Figures 2 and 3. Our model includes five main features: the inner shell, the outer shell, 
the cometary knots, the caps and the high-velocity bipolar outflows.
We modeled the outer and inner shells starting with a sphere and then modified them  to conform with the observed image and {\it  P--V} arrays. For the outer shell we made no effort to deform a sphere since there is no information on the data to guide us in this task,  its real shape may  slightly differ from a sphere but this is of no consequence for the global model and for this reason we call it an spheroid instead of only a sphere. For the inner shell, strings of filaments were also added in its inner sections and over its surface (see Figure 7). 

For the velocity field we started with a homologous expansion, i.e. radial velocity vectors linearly increasing in magnitude with distance and added a poloidal component, as described in \citet{Steffen09} and \citet{lophb5}, however it was found that a velocity field with only radial velocity components  is sufficient to describe the velocity law in this case. The stationary cometary knots where modeled individually from cylinders with emissivity variations along the axis of the cylinder and distributed over a flattened toroid or disk sitting near the equatorial region of the inner shell. For the filamentary and diffuse large scale emission pertaining to the caps we used again sections of spheres with emissivity variations along their axes distributed
over sections of a  flattened torus  placed at the border of the outer shell and projected in front and behind the tilted inner shell. The bipolar collimated outflows were built from long cylinders whose length was arbitrarily limited to the size covered by the set of spectra. 

The results of this process are the synthetic {\it P-V} arrays shown next to the observed ones in Figures 2 and 3. The main uncertainty in the model is the size and the placement of some structures along the line of sight since the information along the third dimension  or into the plane of the sky is derived from the spatially resolved spectra assuming that the nebula is expanding in a homologous way. For example, the bipolar jets may have a larger extent than observed or the caps may be located either at the border of the outer shell, as in NGC 6543, in between the inner and the outer shells or even closer to the inner shell, as in NGC 7009, (see Figure 9). For these reasons our global solution is not  unique in these details, however, since in this case there is a very good spectral coverage and the nebula has so many independent complex elements, the solution is very well constrained by the excellent match of the synthetic {\it  P--V} arrays to the actual data.

The resultant SHAPE mesh model for the Eskimo, rotated 90\degr to the line of sight, before rendering is shown in Figure 7, where the individual main components are labeled. 

The results of the final rendered model are shown in Figure 8, where the top panels show first the synthetic {\it  P--V} array for the central position, slit j, then the synthetic image from the model for the Eskimo as seen on the sky and next to it the same model image but rotated 100\degr{}  into the plane of the sky and also rotated clockwise 45\degr. A composite HST image of the Cat's Eye nebula, NGC 6543 is next presented for comparison.  In the bottom panels we have used the SHAPE model from Steffen et al. (2009) for NGC 7009, the Saturn nebula. That model has been rotated 90\degr{} for this work to show how its  central line profile would appear if we were looking at it pole-on, as in the case of the Eskimo. The corresponding model image is next and to its right is the model image of NGC 7009 as seen on the sky. Finally an RGB {\it HST} image of the Saturn nebula is shown in the last panel for comparison. The structural similarities between the Eskimo, the Cat's Eye and the Saturn nebulae are apparent and striking.

As an additional point of comparison among these PNe, in Figure 9 are presented the observed {\it  P--V} arrays for NGC 6543 and NGC 7009 obtained along their major axes, the data have been drawn from the {\it SPM kinematic catalogue of galactic planetary nebulae} \citep{Lopez12}, together with a synthetic  {\it  P--V} array of the Eskimo obtained also along its major axis after rotating the model nebula 90\degr{} to derive the corresponding bi-dimensional line profile. The velocity scale for the synthetic  {\it  P--V} array has been squeezed by a factor of two and blurred with a gaussian filter to make it comparable with the rest. It is clear that all these  {\it  P--V} arrays (nebulae) share similar structural elements. 

\section{DISCUSSION AND CONCLUSION}

Our aim in this work has been to disentangle in detail the main morphological and kinematic elements of the complex Eskimo nebula, NGC 2392. In order to present the results in an easy to visualize form we have paid particular attention to relate key structures in the projected morphology of the Eskimo from {\it HST} images with spatially resolved, long-slit, echelle spectra. Channel maps and kinematic moment maps have been created from the set of long-slit spectra to aid visualize the kinematic structure of the Eskimo. This relation between morphology and  kinematics has been modeled using the code SHAPE in order to produce a 3D model of the object. We find that many of our results are in general agreement with early works on the Eskimo, cited in the introduction, that gave the first insights into the actual structure of this PN. However, our data and model do not agree with the recent suggestion by \citet{Zhang12} that consider the Eskimo as a twin of Mz-3, a bipolar nebula with a very tight waist and a complex outflow structure external to the main lobes. The present work presents the largest published collection of spatially-resolved, long-slit, echelle spectra for the Eskimo that together with the SHAPE model provide a first, clear, data-based  3D representation of this complex pole-on oriented nebula.

The data  and modeling presented here yields the following results for the different main components of the Eskimo: the outer shell is an oblate spheroid with \vexp{} = 16 \kms. 
This is probably the only structural element that preserves some memory of continuous evolution from the end of the AGB stage up to now. For an outer shell angular radius r = 23\arcsec,  \vexp{}= 16 \kms{} and adopting a distance D = 1.4 kpc, the estimated mean kinematic age would be 9300 years.

The inner shell is a deformed (peanut-like) prolate spheroid with \vexp{} = 120 \kms{} along its major axis. Its orientation is nearly pole-on, tilted $ 9$\degr{} with respect  to the line of sight and pointing towards P. A. = 205\degr{}, i.e. to the south and west. Multiple bright filaments are distributed over its surface that under ground-based seeing conditions give it the appearance of a continuous bright border rim. Its side view shape from the model (see Figures 8 and 9) is very similar to the shape of the inner shells in NGC 6543 and NGC 7009. The velocity law in our model assumes a homologous expansion with a Hubble-type velocity law of the form $v = k \cdot r/r_0$, where $k$ is a constant, $r$ is the distance from the center, and $r_0$ is the distance at which the velocity $k$ is reached. The model does not require a poloidal velocity component in order to match its current expansion conditions and shape, only different values of $k$ are applied to the different components of the nebula . If the outer and inner shell formed at approximately the same time, at the current expansion rate of the latter along its major axis  it would  have overtaken the outer shell in only 1250 years. Therefore, either the inner shell had a much slower acceleration in the past or its present expansion rate has been in operation for a relatively short time compared to the development of the outer shell. Although the current exact separation between the approaching tip of the inner shell and the border of the outer shell cannot be discerned, our model assumes that the inner shell has not reached the border of the outer shell yet. The fast expansion of the inner shell along a preferred symmetry axis occurred at a much later time than the exit from the AGB stage, once the envelope was fully ionized and the central star had developed a sufficiently fast wind to influence the expansion of the shell through the mechanical and thermal energy produced by the shocked wind on the previously expelled matter. Once the fast expansion of the inner shell started, this clearly took a non-isotropic mode. The inner shell harbors a hot bubble with extended X-ray emission but \citet{Guerrero05} point out that the stellar wind velocity does not seem to be high enough to account for the X-ray luminosity and briefly discuss the possible contributions from a binary companion and/or  the fast bipolar outflow to resolve this discrepancy.

The bipolar jet is detected emerging right from the star at \vexp{} = 165 \kms, it has a near pole-on orientation, with its axis tilted some 5\degr{} further to the southwest than the major axis of the inner shell; the jets get split into two sections as they leave the inner shell and then reach  \vexp{} = 180 \kms, the jets show an approximately constant width all along its detected length, indicating a high degree of collimation. Its projected length is 25\arcsec{} but tilted some 30\degr  with respect to the line of sight, assuming a constant velocity, the jets have been expanding for approximately  1800 years, without considering possible episodic events. 

The caps are distributed in three main extended sections that in projection form most of the fur of the Eskimo's hood. Each of these sections has a bright portion that is conspicuous in the fur of the Eskimo's hood, but the real extent of each of them is larger than the bright portions as indicated by the spectroscopic information. Two of these sections are in the back part, mainly to the north and west of the inner shell and show the corresponding receding velocities, the third section is  in the front and to  the south of the inner shell and shows the expected approaching velocity; the caps are expanding at \vexp{} = 55 \kms.  These caps correspond to what \citet{Odell85} called wisps and later \citet{Odell02} called the north, west and south fuzz. The caps are not aligned with the axis of
the bipolar jet, a similar situation occurs in NGC 6543 (see Figure 9). Considering that the inner shell expands twice as fast as the caps and these seem to be located beyond the tips of the  inner shell,  this suggests that either the caps (perhaps together with the equatorial disk of cometary knots) were part of a pulse of mass loss that occurred sometime after the formation of the PN or are part of the outer shell. In our model the caps are placed at the edge of the outer shell, if their expansion velocity has been constant through time, it would take them 2700 years to reach the outer envelope from the star and, as mentioned before, the inner shell would have caught up with them in half that time. The reasonable alternative is then that they are originally part of the outer envelope. The structure of the caps in our model is very similar to the caps in NGC 6543 (see Figure 9).

The cometary knots are distributed over an equatorial disk, external to the inner shell at a radius of $\sim 17\arcsec$ that projects them over the fur of the Eskimo's hood and mixes them with the caps; these  knots are at the systemic velocity, i.e. they seem inert or with no radial velocity component, it is likely that they are slowly expanding radially with respect to the central star and have been formed close to its current location, or near the main ionization front, as discussed by \citet{Odell02}. Note that a similar equatorial  knotty ring is present in NGC 6543 (see Figure 9) and traces of a similar structure are discernible in NGC 7009.
 
Our model has lead us in a natural way to compare the striking similarities of the Eskimo nebula with NGC 6543 (the Cat's Eye) and NGC 7009 (the Saturn) nebulae, and by extension with the rest of the group of elliptical nebulae with ansae or FLIERS (e.g. NGC 3242, NGC 6826 and NGC 7662). \citet{Balick98} have described in detail the characteristics of these PNe, they all show a closed inner shell with a bright rim, an outer shell with low ionization emission regions and a collimated bipolar outflow. Comparing the images in Figure 8 it is straightforward to relate the curious, peanut-like, shape of the inner shell in the Eskimo with the inner shells in NGC 6543 and NGC 7009. An equatorial disk of cometary knots as the one described for the Eskimo is apparent in the image of NGC 6543 and the structures related to the caps in the Eskimo can also be clearly identified in NGC 6543 and NGC 7009, though in the latter they seem to be distributed over a cylindrical shell right outside the inner shell.  In both cases they also have bipolar collimated outflows. With these analogies in mind it becomes easy to visualize how the Cat's Eye nebula and the Saturn nebula would look like if viewed pole-on, very much like the Eskimo. \citet{Steffen09} have made a model of the Saturn nebula and its pole-on view is shown in the second frame of the lower panel of Figure 9. 

This group of elliptical nebulae with ansae or FLIERS is particularly interesting since as \citet{Kastner12} show,  all the members of this group are among the relative small number of PNe where extended X-ray emission has been detected. In addition, they all lie at distances from the galactic plane that range from $\simeq$ 450 -- 800 pc, exceeding the PN scale-height of 250 pc for elliptical PN. This suggests that the progenitors are old-disk stars of relatively low mass but still massive enough so their post-AGB wind evolve at an adequate pace to be able to develop a closed inner shell and a hot bubble. If the remnant stellar mass is too low, the stellar wind will accelerate at a very slow rate, keeping the initial two-wind interaction in a momentum conserving state and the shocked wind region remains isothermal, allowing the shell to evolve and expand, thus never reaching conditions to form a hot, X-ray emitting bubble.

Although no binary companion has been found in the central star of the Eskimo, there are strong arguments in favor of its existence. The X-ray, extended and point, emission \citep{Guerrero05, Kastner12} cannot be explained by the the relatively low stellar wind velocity, $V_{\infty} \simeq 400$ \kms {} and low effective temperature T$_{\rm eff} \simeq 45 \times 10^3$ K of the central star; the star has a mass loss rate \.{M} $\leq 3.2 \times 10^7$ M$_{\odot}$~yr$^{-1}$,  \citep{Kasch12}. The characteristics of the central star of NGC 2392 are also inconsistent with the high stages of ionization observed in the nebula or the significant Zanstra discrepancy derived from the photoionization models \citep[e.g.][]{Danehkar12}. 

 Furthermore, the high expansion velocity of the inner shell  seems to require a driving force beyond the modest characteristics of the central star since it is only the slightly higher thermal pressure of the hot X-ray emitting gas over that of the inner shell that drives its expansion \citep{Guerrero05}, and there is still the fast  bipolar collimated outflow that needs a plausible launching engine. All these elements favor the presence of a binary companion to the central star of NGC 2392.
  
Just as in the case of the Eskimo, there is no firm evidence of binary nuclei in the other members of the group of elliptical nebulae with ansae or fliers. Recently, \citet{Douchin12} list NGC 6826 as a possible binary with a period of 0.619 days, but \citet{Jevtic12} after a year of  continuos observing with Kepler find only stochastic brightness variations in the central star of NGC 6826. \citet{Prinja12} consider that  these brightness variations found by \citet{Jevtic12} in the central star of NGC 6828 are possibly related to subsurface convective layers, just as the ones they have found in the central star of NGC 6543 and where they derive a high rotational velocity for the central star. High stellar rotation in the central star of a PN may be an indication of spin-up by a  past merger, tidal interaction or mass accretion in a binary system. High rotational stellar velocities help also in the production of toroidal magnetic fields that serve as launching platforms for bipolar jets \citep[e.g.][]{Gar05}

A detailed reexamination of the evolutionary history, the possible common mechanisms and sequence of physical events in the formation of  the members of this group that make them share so many outstanding characteristics is evidently called for, but out of the scope of the present paper. 

We have presented a detailed morpho-kinematic analysis of the Eskimo planetary nebula, NGC 2392. Our resultant 3D model reveals that this nebula is a close analog to the Cat's Eye (NGC 6543) and the Saturn (NGC 7009) nebulae once the orientation effect is considered.

This research has benefited from the financial support of DGAPA-UNAM
through grants IN116908, IN108506, IN100410, IN110011 and CONACYT 82066.  
We acknowledge the excellent support of the technical personnel at the OAN-SPM.
We thank the reviewer of this work, Dr. D. Frew, for his constructive comments that helped
to improve the presentation of this paper.

\vfill \eject

\vfill \eject

\begin{figure*}[!t]
    \centering
  \includegraphics[width=0.8\textwidth]{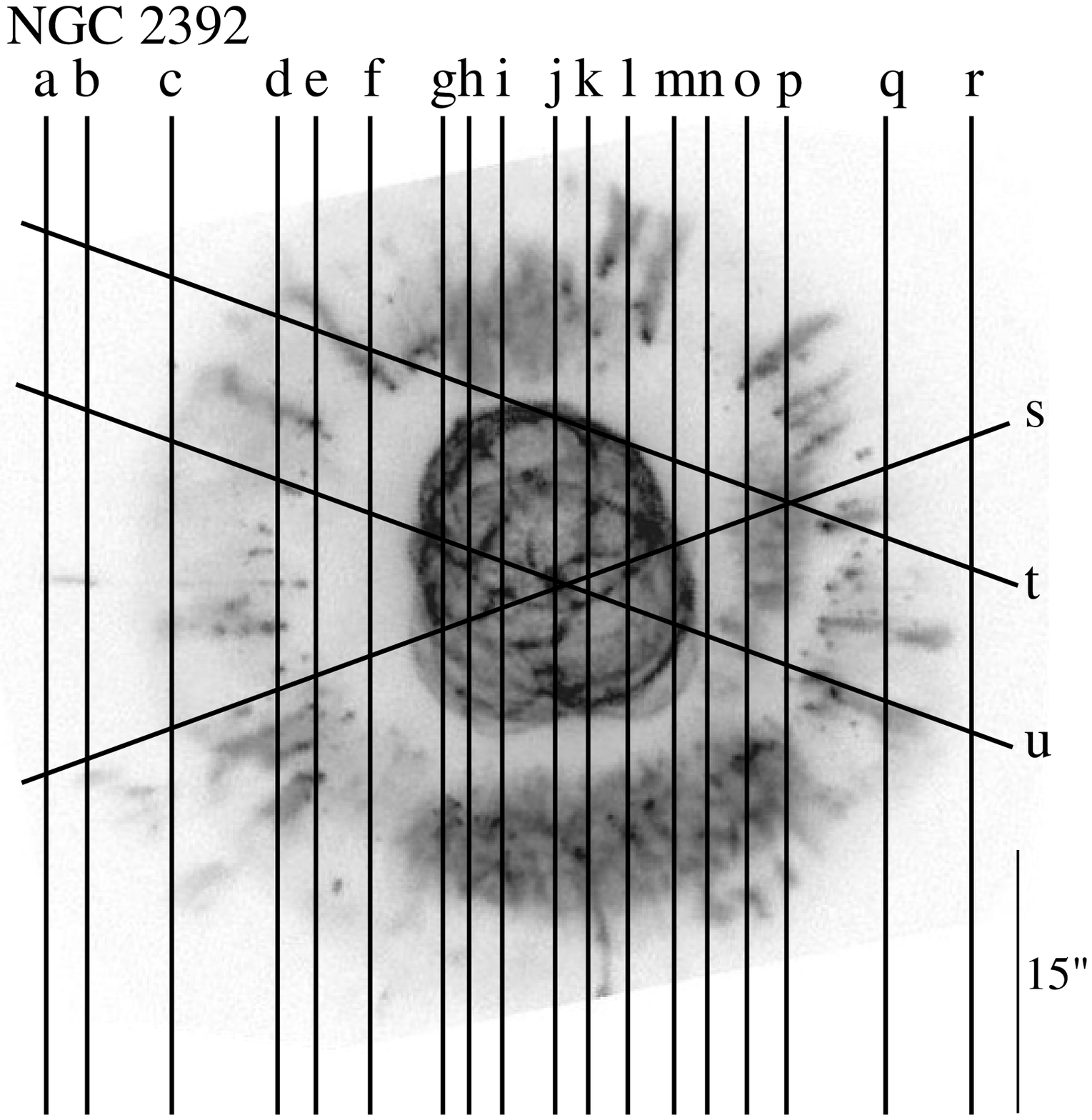}
  \caption{Location of each slit position is indicated and labeled on an
    HST \NIIlam{} image of the Eskimo. North is up, east left.}
\end{figure*}

\begin{figure*}[!t]
\centering
  \includegraphics[width=0.7\textwidth]{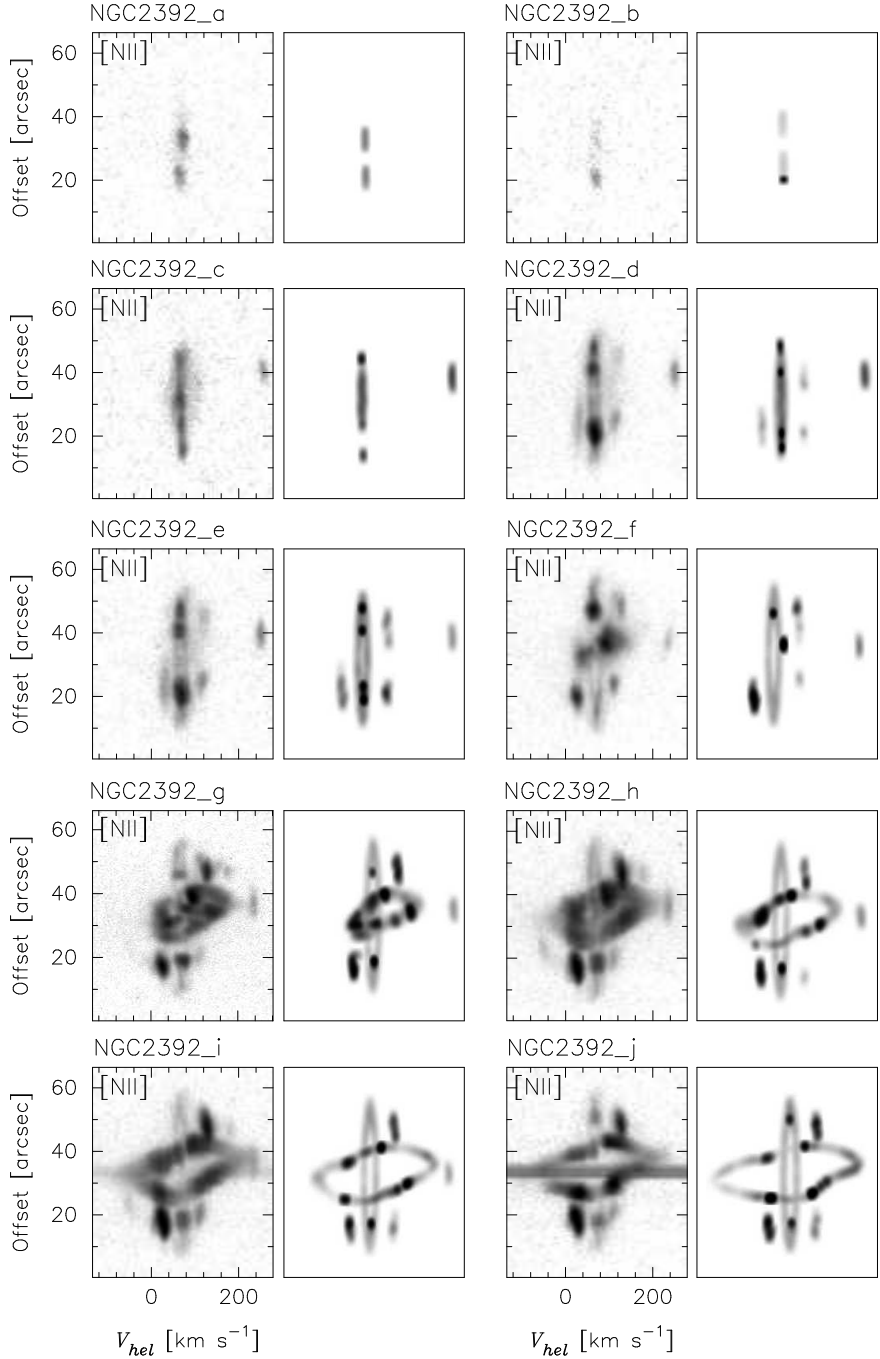}
  \caption{Mosaic of bi-dimensional {\it  P--V} arrays labeled according to slit position. For each slit position we show a couple of {\it  P--V} arrays: the observed \NIIlam{} {\it  P--V} array is on the left panel and the corresponding synthetic {\it  P--V} array derived from the model is on the right panel. This figure shows slit positions a-j, the remainder positions are shown in Figure 3. The top of the array corresponds to the north end of the slit.}

\end{figure*}

\begin{figure*}[!t]
\centering
  \includegraphics[width=0.7\textwidth]{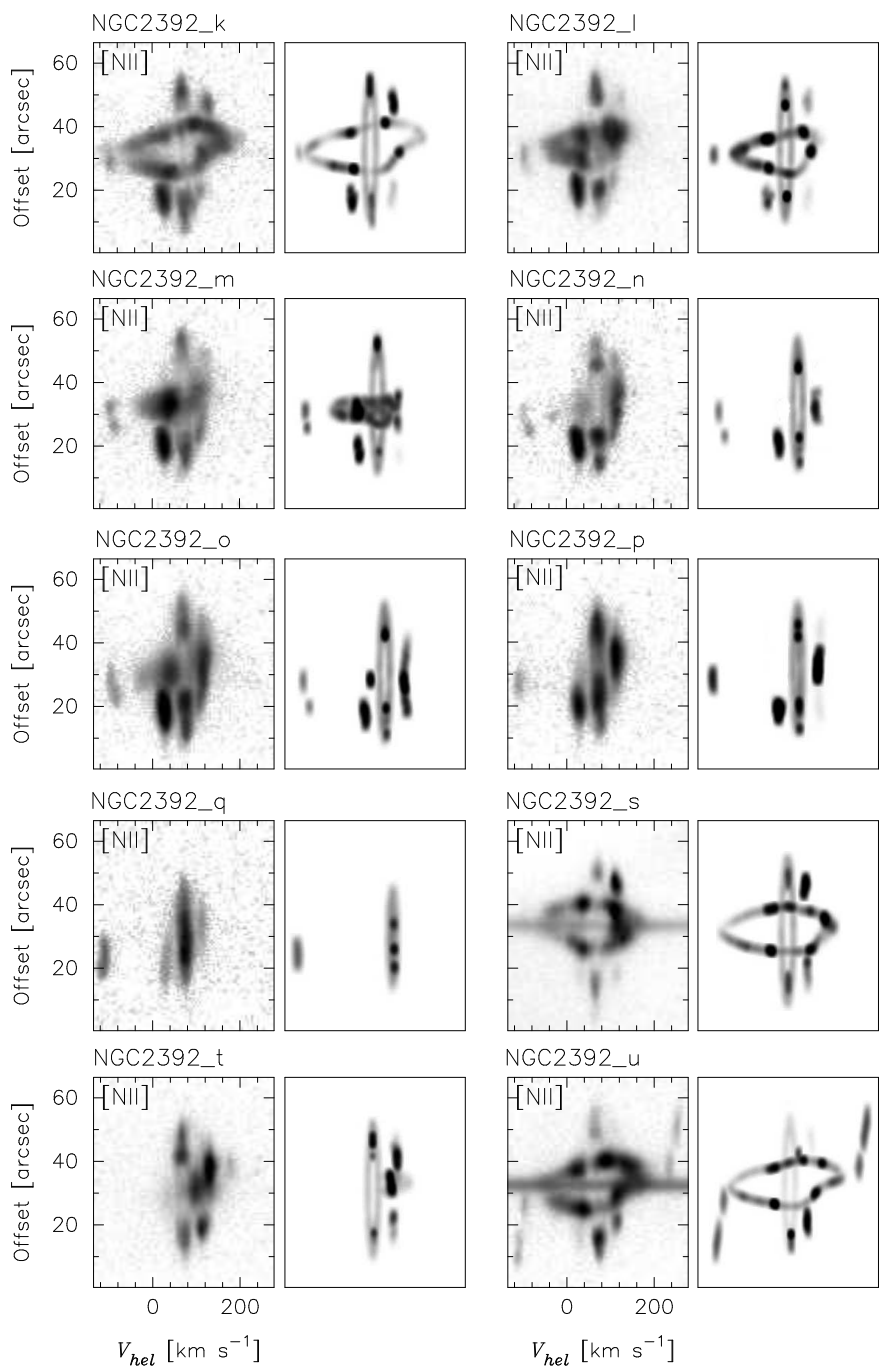}
  \caption{As in Figure 2 but for the slit positions k - u. The top of the array corresponds to the north end of the slit for slit positions k - q. For slit position s the top of the array corresponds to the northwest end of the slit and for slit position t and u the top of the array to the northeast end of the slit. }
\end{figure*}

\begin{figure*}[!t]
\centering
  \includegraphics[width=0.8\textwidth]{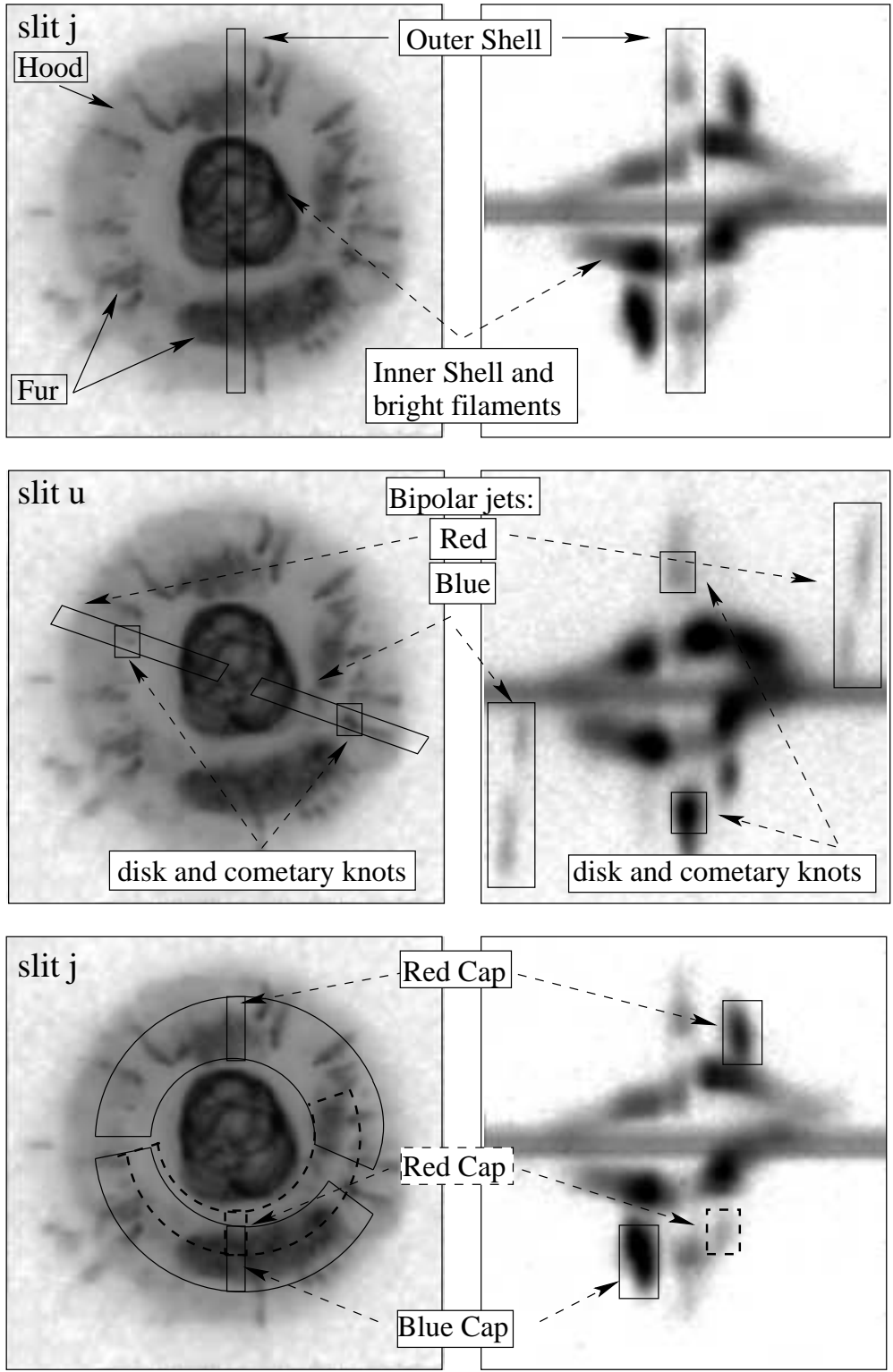}
  \caption{Finding chart for individual emission regions seen in the {\it  P--V} arrays in slit positions  j and u. Each of these components is present  throughout the spectra presented in Figures 2 and 3
  }
\end{figure*}

\begin{figure*}[!t]
\centering
\includegraphics[width=0.85\textwidth]{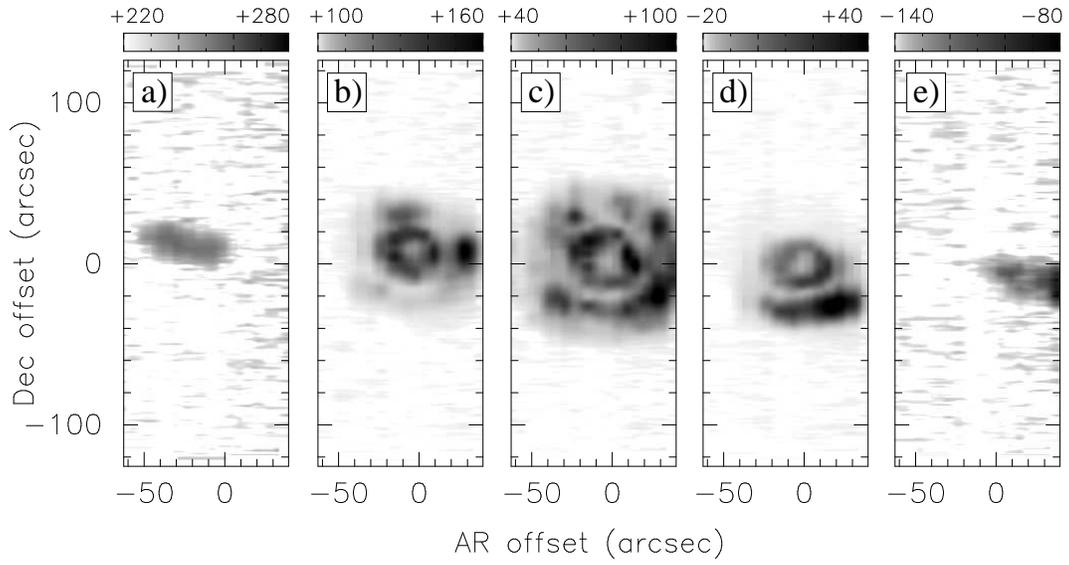}
  \caption{\NIIlam{} emission-line channel maps constructed from the
    long-slit echelle spectroscopy. Each isovelocity image is 60
    \kms{} wide.
    }
\end{figure*}

\begin{figure*}[!t]
\centering
  \includegraphics[width=0.85\textwidth]{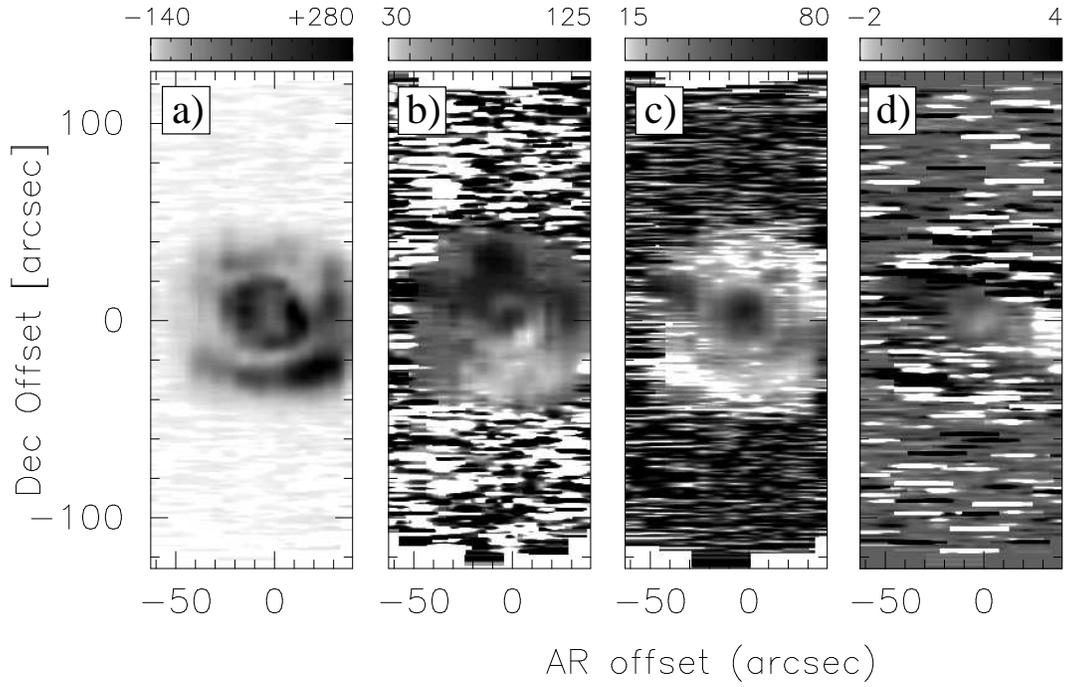}
  \caption{Moment maps for the \NIIlam line. Panel a) Surface brigthness;
    b)  mean heliocentric velocity; c)  rms velocity and d) skewness. All moment maps are integrated over the full velocity range, \vhel=$-$140 to $+$280~\kms, but the scale bars at the top of the maps show only the interval for the most representative values displayed within the nebula, excluding the background}
\end{figure*}

\begin{figure*}[!t]
\centering
  \includegraphics[width=0.8\textwidth]{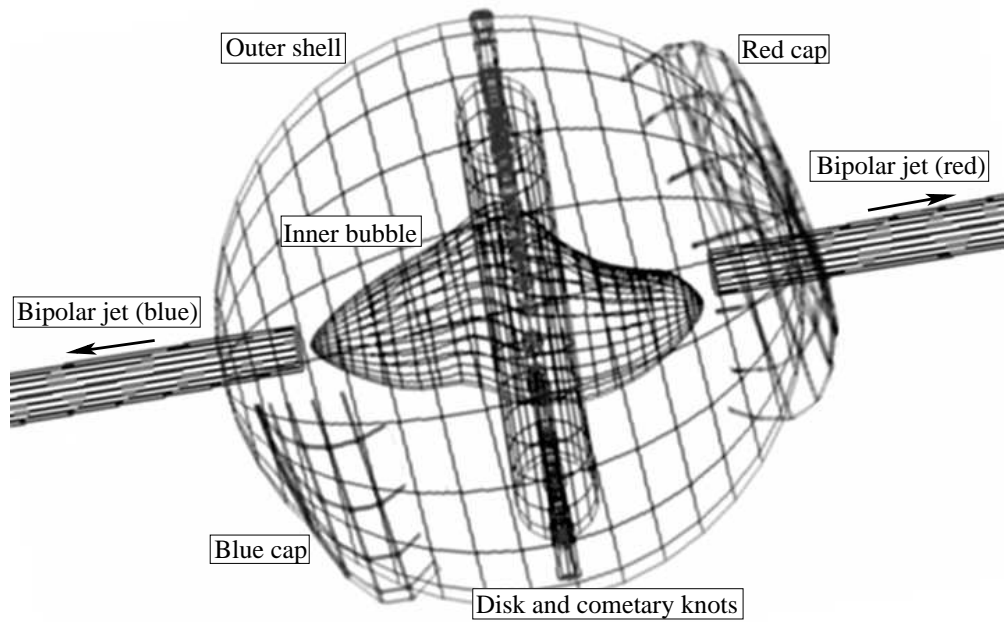}
  \caption{SHAPE mesh model of the Eskimo, rotated 90\degr{} to the line of sight, before rendering. 
  The individual main components are labeled.}
\end{figure*}

\begin{figure*}[!t]
\centering
  \includegraphics[width=0.85\textwidth]{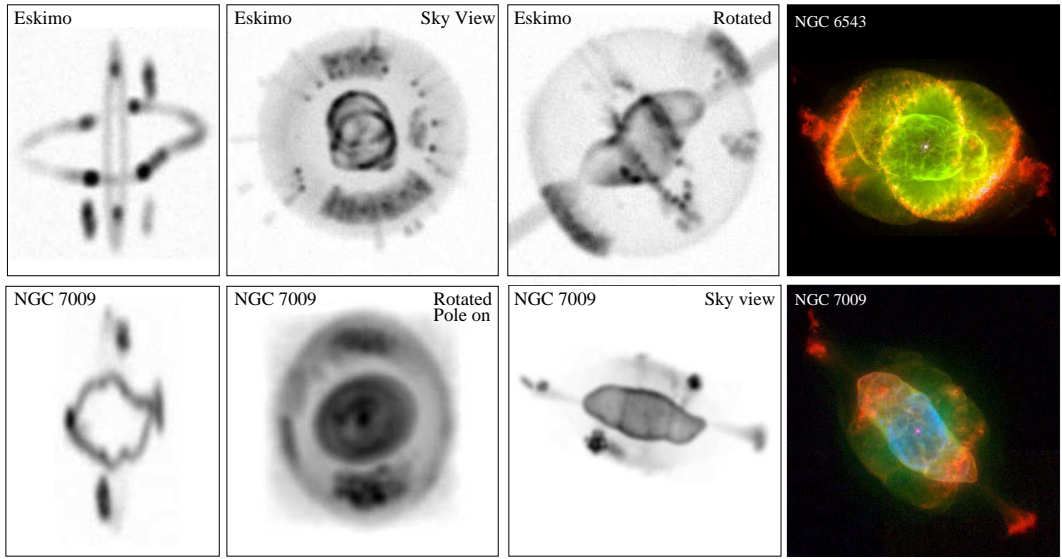}
  \caption{ Upper panels, from left, The synthetic line profile for the
    central slit j. The Eskimo image, as seen on the sky,
    derived from the model. The previous image rotated 100\degr, into the plane of the sky and 45\degr{} clockwise. A composite Chandra + {\it HST} image of NGC 6543, shown for comparison.
      Lower panels:   A synthetic line profile from a slit located at the center of a
    pole-on view of NGC 7009, from the SHAPE model of Steffen (2009). The corresponding model image of NGC 7009 as it would appear if seen pole-on. The model  image of NGC 7009
   as seen on the sky. {\it HST} image of NGC 7009 shown for comparison. Image credits for the {\it HST} images: NGC 6543, J. P. Harrington \& K. J. Borkowski. NGC 7009, B. Balick, NASA/HST. }
\label{fig:mosaico}
\end{figure*}

\begin{figure*}[!t]
\centering
  \includegraphics[width=0.55\textwidth]{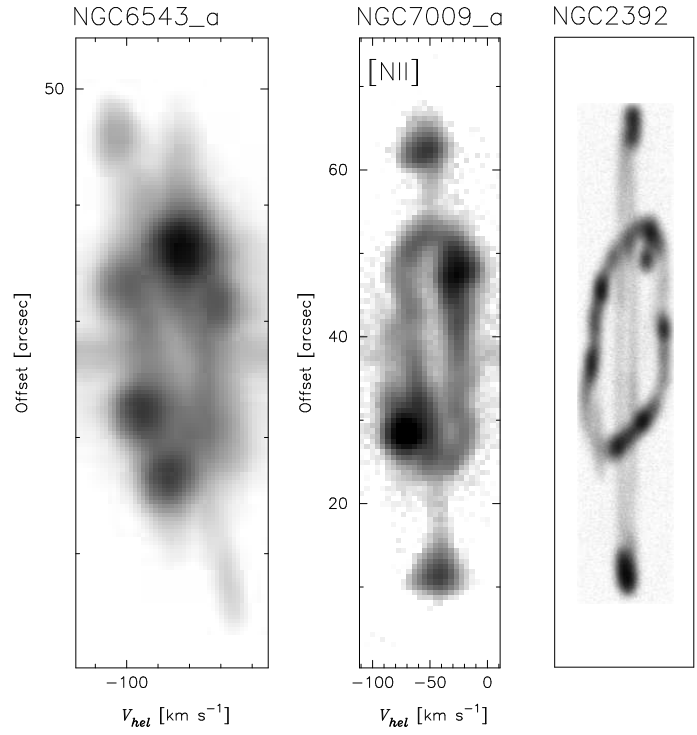}
  \caption{Left and central panels, the observed line profiles for a
    central slit in NGC 6543 and NGC 7009, respectively. Right panel, the synthetic line profile for NGC 2392 showing how it would like along the axis from a side view, as the former cases. 
    }
\end{figure*}

\end{document}